# Imbibition of Oil in Dry and Prewetted Calcite Nanopores

Ejaz Ahmed,[1] Huajie Zhang,[1] Mert Aybar,[1] Bikai Jin,[2] Shihao Wang,[2] and Rui Qiao[1,*]

[1] Department of Mechanical Engineering, Virginia Tech, Blacksburg, VA 24061, USA
[2] Chevron Energy Technology Co., 1500 Louisiana Street, Houston, TX 77002, USA

**Abstract.** Fluid imbibition into porous media featuring nanopores is ubiquitous in applications such as oil recovery from unconventional reservoirs and material processing. While the imbibition of pure fluids has been extensively studied, the imbibition of fluid mixture is little explored. Here we report the molecular dynamics study of the imbibition of model crude oil into nanometer-wide mineral pores, both when pore walls are dry and prewetted by a residual water film. Results show the fastest imbibition and quickest propagation of molecularly thin precursor films ahead of the oil meniscus in the dry pore system. The presence of a thin water film on pore walls corresponding to an environmental relative humidity of 30% slows down but still allows the spontaneous imbibition of single-component oil. Introducing polar components into the oil slows down the imbibition into dry nanopores, due partly to the clogging of the pore entrance. Strong selectivity toward nonpolar oil is evident. The slowdown of imbibition by polar oil is less significant in the prewetted pores than in dry pores, but the selectivity toward nonpolar oil remains strong.

---

[*] To whom correspondence should be addressed. Email: ruiqiao@vt.edu.

# 1. Introduction

The imbibition of fluids into porous media with nanoscale pores plays a key role in applications such as oil recovery from unconventional reservoirs dominated by nanopores, filtation in clay, and manufacturing of ceramic matrix composites.[1-6] Understanding the thermodynamics and kinetics of fluid imbibition is crucial for optimizing these applications and thus has been pursued using experimental, theoretical, and computational methods. Among the available works, those focusing on the imbibition at the pore scale are foundational in that they provide the basis for understanding and predicting the imbibition behavior in porous media.

Molecular dynamics (MD) simulations have been widely used to study fluids imbibition into nanopores because they allow molecular details of fluids and fluid-wall interactions to be resolved with relatively few assumptions. Early work revealed that even in nanopores with a width of a few tens of fluid diameters, the imbibition rate can follow the diffusive scaling law predicted by capillary flow theories, e.g., the propagation length of the imbibition front is proportional to the square root of time.[7] Deviations from the predictions from classical capillary flows have also been reported, e.g., due to slippage at liquid-wall interfaces.[8-10] Furthermore, it was shown that other mechanisms such as surface hydration can lead to imbibition. For example, when the capillary flow is blocked by gas in a strongly hydrophilic nanopore, water can still imbibe into the pore in the form of molecularly thin films and the imbibition rate also follows the diffusive scaling law.[11]

Recent pore-scale work on imbibition in nanopores focused on more complex situations than earlier works. For example, the imbibition of water into hydrophilic nanopores initially filled with oil has been studied widely due to its relevance to oil recovery from unconventional reservoirs.[12-14] These works identified the essential role of fluid-surface interactions in imbibition. For example, strong interactions between hydrocarbons and inorganic pore walls can hinder the displacement of oil by water during spontaneous imbibition.[15] Pore shape (or more fundamentally, the curvature of pore surfaces), wall defects, and heterogeneous wettability of pore walls, which modify local fluid-surface interactions, have also been found to affect imbibition kinetics.[16-21] For example, Xie and colleagues studied the detachment of oil molecules from defective alumina surfaces driven by the spontaneous imbibition of water and showed that surface defects provide additional adsorption sites but also facilitate the detachment of oil molecules in aqueous solutions.[19]

Previous pore-scale studies have advanced our understanding of imbibition in nanopores, in particular, the role of fluid-surface interactions in regulating the imbibition. However, these studies have focused on relatively idealized scenarios as far as the fluids in their systems are concerned. First, they usually dealt with pure fluids, and thus how the multi-component nature of imbibed fluids (or the presence of impurities in them) affects imbibition received only limited attention.[17, 22, 23] Second, the presence of impurity fluids on pore walls before imbibition is rarely considered. For high-energy mineral surfaces in the natural environment, molecularly thin water films often develop on them.[24] However, how the prewetting of pore walls by these films affects imbibition has rarely been studied.

In this work, we study the imbibition of pure and mixture oils into dry and prewetted nanopores using MD simulations. We seek to delineate how the impurities in the imbibed fluids and water initially on pore walls affect imbibition and elucidate the role of surface-fluid interactions in these effects. The rest of the manuscript is organized as follows. Section 2 presents four systems studied and the models and methods adopted. Section 3 presents the qualitative picture and quantitative dynamics of imbibition in the four systems considered and discusses the underlying mechanisms. Finally, conclusions are drawn in Section 4.

## 2. Simulation systems, protocol, and methods

### 2.1 Simulation systems

In this study, we investigated the spontaneous imbibition of model crude oil into a single calcite nanopore. As illustrated in Figure 1, the simulation system consists of a slit-shaped calcite pore, blocker atoms sealing the pore initially, and an oil reservoir. The calcite pore, aligned along the *x*-direction, is open at the left end and connected to the oil reservoir at the right end. The oil reservoir was exposed to free space so that its pressure was close to zero. Our simulations were typically terminated when the oil reached the pore's left end. This way, imbibition is not notably obstructed by the fluids ahead of the capillary imbibition front nor affected by external forces, which mimic spontaneous fluid imbibition into a long pore with negligible initial *bulk* fluids. The imbibition in the systems studied here resembles one of the key steps of the laboratory characterization of core samples from tight reservoirs. Specifically, core samples with original hydrocarbons removed are often placed in hydrocarbon under a vacuum to saturate the sample for subsequent testing (see, e.g., Ref. [25]).

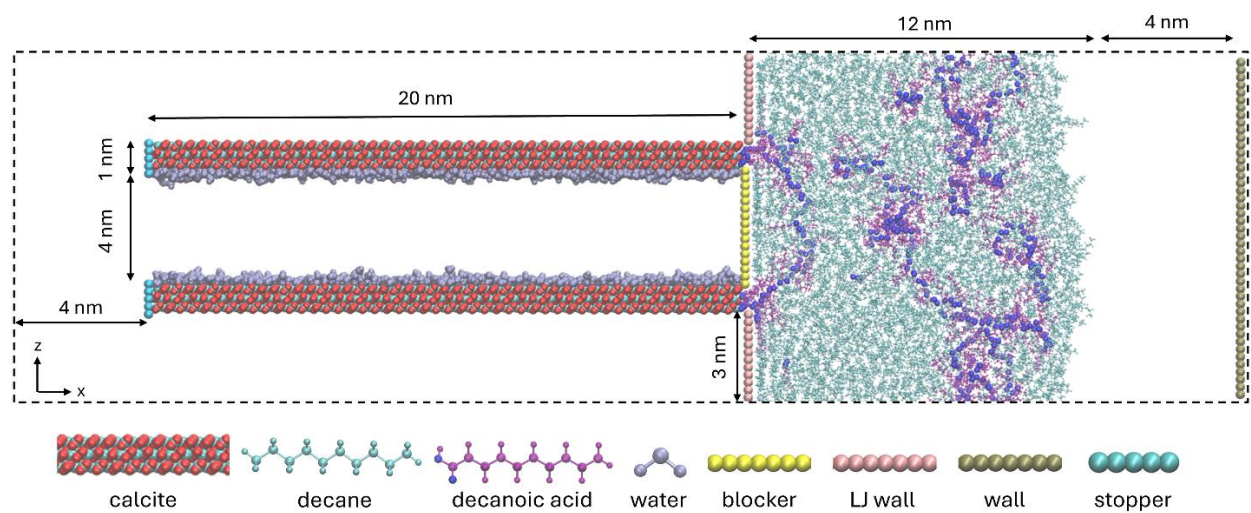

**Figure 1**. A snapshot of the molecular system for studying the imbibition of fluids (here, a 3:1 decane-decanoic acid mixture) into a calcite nanopore with walls prewetted by molecularly thin water films. The dashed black line denotes the periodic simulation box.

To capture nanoscale confinement effects on imbibition, the pore width was set to 4 nm. This dimension is representative of pore sizes commonly found in tight formations and shale reservoirs. [26, 27] The pore length was set to 20 nm so that the aspect ratio of the pore is much larger than 1.0, as in natural nanoporous networks. Further, a pore of this length provides sufficient space to observe the progression of the imbibition front and ensures that the underlying physical mechanisms can be thoroughly examined through MD simulations.

Four systems were considered to study how oil compositions and prewetting of the nanopore affect imbibition. In the first two systems, initially, the calcite pore was free of any fluid molecules, and the reservoir was filled with pure n-decane ($C_{10}H_{22}$) or a mixture of decane and decanoic acid ($C_{10}H_{20}O_2$) at a mole ratio of 3:1. We note that pure decane, a nonpolar fluid, is widely used as a model oil in MD simulations. On the other hand, introducing amphiphilic decanoic acid into oil allows it to more closely resemble real crude oil (the mole fraction of polar oil used here is higher than in typical crude oil. However, using a much smaller polar oil fraction requires a much larger number of nonpolar oil molecules to be simulated and increases the statistical uncertainty of polar component imbibition, even though it would not change the imbibition behavior qualitatively. Therefore, a relatively high mole fraction is used for polar oil). The comparison of the imbibition in these systems (hereafter referred to as dry pore systems) helps isolate the role of polar motifs in oil molecules − such as decanoic acid – in the imbibition process. In the other two systems (hereafter referred to as prewetted pore systems), initially, a molecularly thin layer of water molecules was introduced on the calcite pore walls while two types of oil (pure decane and 3:1 mixture of decane with decanoic acid) were again considered. The amount of initial water on pore walls corresponds to calcite walls in equilibrium with an environment with a relative humidity of 30%, which is within the range encountered in shale reservoirs or laboratory imbibition testing of core samples. At this initial water loading, water molecules form a thin film on each wall with an area density of 11.06 nm$^{-2}$ (see below), which leads to a water saturation of ~ 0.17 in the 4 nm-wide pore studied here. The prewetted pore systems mimic laboratory testing of core samples that are not completely dried, which is often desired because drying process can potentially alter the pore network and kerogen structure.[25]

For all systems, the simulation box is approximately 40.9 nm in the *x*-direction (the pore axis direction), 2.5 nm in the *y*-direction, and 11.7 nm in the *z*-direction. Periodic boundary conditions were used in all three directions. The relatively small *y*-dimension is deliberately chosen to balance computational efficiency with the need to minimize finite-size effects and ensure statistical representativeness of the molecular configurations.

## 2.2 Simulation protocol

The simulation of each system described above consisted of two steps. In the first step, blocker atoms (Fig. 1) were placed at the pore entrance to physically prevent oil molecules from entering the nanopore and an equilibrium run of 2 ns was performed. This was followed by another 18 ns of simulation during which 4-5 system configurations were saved. In the second step, the simulation started from the saved configurations in the first step, and the blocker atoms were removed (defined as $t = 0$) to initiate the imbibition process. The systems were then evolved 20 ns and 50 ns for the dry and prewetted pore systems, respectively. These production timescales were selected to ensure that the imbibition front could propagate sufficiently into the pore, allowing observation of flow behavior and molecular structuring. To ensure statistically robust results, the imbibition results obtained from the different initial configurations were averaged.

During all simulations, the calcite wall and Lennard-Jones (LJ) wall atoms in Figure 1 were fixed. Additionally, for the prewetted systems, stationary "stopper" atoms were placed at the pore's left end. These stopper atoms prevented the water layer from spilling out or redistributing away from the pore interior, thereby preserving a stable, controlled wettability state at the pore surface. Note that, as in the dry pore systems, the left end of the calcite pore is open to free space, which has a negligible pressure.

**2.3 Molecular models**

The calcite pore walls were modeled as rigid slabs, 1 nm in thickness, oriented along the [1014] crystallographic plane. The atomic partial charges and LJ parameters of calcite walls were derived from the refitted Doves' potential.[28] The OPLS-AA (Optimized Potentials for Liquid Simulations– All Atom) force field was employed to describe both decane ($C_{10}H_{22}$) and decanoic acid ($C_{10}H_{20}O_2$) and was generated using the LigParGen server.[29-31] The thin water film coating the pore surface, present in the prewetted pore systems, was represented by the extended simple point charge (SPC/E) water molecules.[32] The neutral "blocker", "LJ wall", "wall" and "stopper" atoms in Figure 1 were modeled as LJ atoms arranged on a square lattice. Because their main purpose is to form a physical boundary, their LJ parameters were selected to ensure that the adhesion of fluid molecules, particularly water, is weak. A listing of LJ parameters for calcite, hydrocarbon, acid, water molecules, and auxiliary components such as LJ wall and stopper, can be found in Table S1 of the Supplemental Online Material.

**2.4 Simulation methods**

All MD simulations were performed using the GROMACS code.[33] Simulations were carried out in the NVT ensemble. The temperature of fluids was maintained at 298 K using the velocity rescaling thermostat,[34] which provides smooth temperature coupling. Short-range, non-bonded interactions (e.g., LJ potentials) were computed using the cut-off scheme (cutoff distance: 1.1 nm). For electrostatic interactions, the Particle Mesh Ewald (PME) method was employed with a real space cutoff of 1.1 nm and an FFT spacing of 0.12 nm, thus ensuring accurate representation of coulombic forces for charged and polar species. Neighbor lists were constructed using the Verlet buffer scheme with a 1.1 nm cut-off length. The time step for numerical integration of the equations of motion was set to 1.5 fs in all simulations.

In the prewetted pore systems, the area density of water on each calcite pore wall before imbibition started was computed via separate grand canonical Monte Carlo (GCMC) simulations. Specifically, water molecules were injected into or deleted from a 4 nm-wide slit calcite pore that is periodical in the lateral directions to ensure that the partial pressure (and thus chemical potential) of water in the pore matches that of water vapor at 298 K and relative humidity of 30%. The GCMC runs were performed using the Lammps code through the "fix gcmc" module.[35] GCMC operations were done every 100 MD steps and an average of 10 exchange attempts were conducted. The area density of water on each calcite wall reached an equilibrium value of 11.06 $nm^{-2}$ in 2 ns.

## 3. Results and discussion

**3.1 Imbibition in dry nanopores**

In this section, we examine the imbibition of pure decane and a 3:1 mixture of decane and decanoic acid into dry calcite nanopores. Figure 2a shows the evolution of the total number of oil (decane and decanoic acid) molecules ($N$) in the pore. The spontaneous imbibition is consistent with the complete wetting, and hence significant capillary pressure, of light oil such as hexane on calcite surfaces.[36] The imbibition of pure decane is much faster than that of decane+decanoic acid mixture. However, the imbibition of both fluids shows a linear increase of $N$ as $t^{1/2}$, consistent with the square-root time dependence of fluid penetration in ideal capillaries predicted by classical capillary imbibition theories when initial effects are negligible.[37] For the imbibition of decane and decane+decanoic acid mixture, we introduce a selectivity factor $S(t)$ to quantify the preferential intake of oil components:

$$S(t) = \frac{n_{da}^p(t)}{n_d^p(t)} \times \frac{n_d^r(0)}{n_{da}^r(0)} \tag{1}$$

where $n_{da}^p(t)$ and $n_d^p(t)$ are the number of decanoic acid and decane molecules inside the pore at time $t$, respectively. $n_d^r(0)$ and $n_{da}^r(0)$ are the initial number of decane and decanoic acid molecules in the oil reservoir. $S(t) = 1$ indicates that there is no preferential imbibition of either component. $S(t) < 1$ corresponds to the preferential imbibition of decane over decanoic acid.

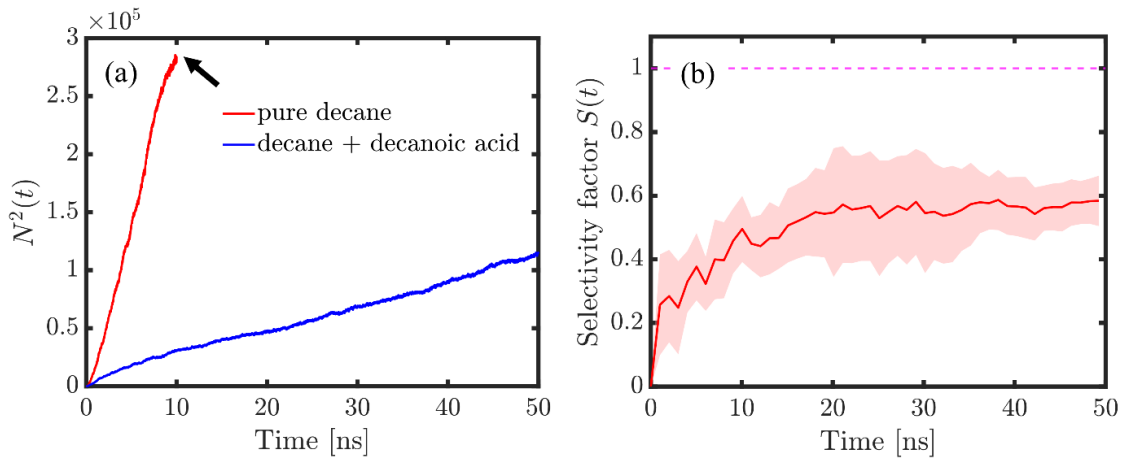

**Figure 2**. The evolution of the total number of oil molecules (decane + decanoic acid) imbibed into the dry calcite nanopore (a) and the selectivity of decanoic acid relative to decane (b) as a function of time. The arrow in (a) marks the instant when the meniscus front reaches the pore's open end. The shaded area in (b) shows the standard deviation of the selectivity factor.

Figure 2b shows the evolution of the selectivity factor $S(t)$ with time for the imbibition of decane+decanoic acid mixture. $S(t)$ is close to 0.2 initially and increases to ~0.6 at t = 20 ns, and varies a little later. These data indicate a preferential uptake of the nonpolar decane molecules relative to decanoic acid through imbibition. Therefore, for core samples from unconventional reservoirs dominated by inorganic nanopores (e.g., core samples from the Niobrara shales),[38] after the initial oil in them is fully removed, saturating them in crude oil through imbibition will likely cause the oil inside the core sample to be enriched in nonpolar components compared to the crude oil.

To elucidate the mechanism governing the above oil imbibition behaviors, we first examine the molecular picture of the imbibition process (Fig. 3). In the imbibition of decane (panels a-1 and a-2), as expected from capillary theories, a liquid meniscus is observed to propagate toward the nanopore's left end. Further, a molecularly thin decane film is observed to move ahead of the meniscus. Such a thin film, typically referred to as the precursor film in the droplet spreading literature,[39] has also been observed in the imbibition of fluids into nanoporous solids.[4] It originates from the far stronger affinity of decane molecules to the calcite walls than to decane molecules and is known to facilitate the recovery of decane from dead-end calcite pores.[40] To appreciate the strength of the decane-calcite wall interactions relative to the decane-decane interactions, we computed the interaction energy between the contact-adsorbed $CH_2$ motifs of decane molecules with the calcite wall (a motif is considered contact adsorbed if it is located within 0.5 nm i.e., the first absorption layer of the topmost atom layer of the calcite wall). Table 1 shows that the $CH_2$-calcite interaction energy is -6.02 kJ/mol, which is almost 8 times stronger than the interaction energy of $CH_2$ with decane molecules in the bulk (-0.75 kJ/mol). It is thus energetically favorable for a decane molecule to move from the bulk-like meniscus to the calcite surface to drive the formation of the precursor film.

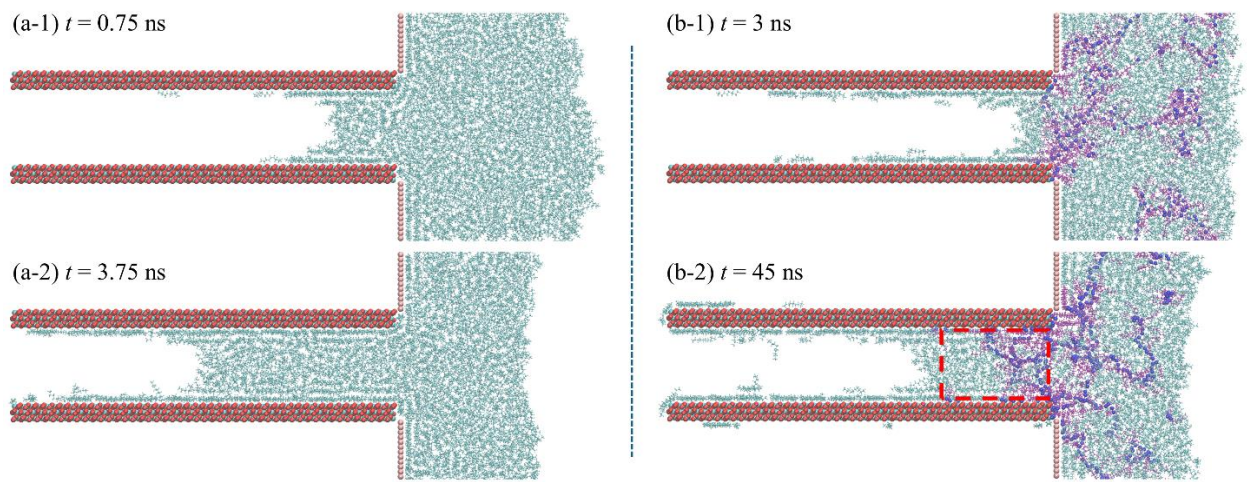

**Figure 3**. Snapshots of the imbibition of pure decane (a1-a2) and decane-decanoic acid (b1-b2) into a dry calcite nanopore. The color coding and molecular representation are the same as that in Fig. 1.

The ready formation of the precursor film, in which the decane molecules maintain mobility, provides an efficient pathway for the ingression of decane into the calcite nanopore. This pathway, along with the capillary flow, contributes to the rapid imbibition of pure decane into the calcite pore.

Figures 3b-1 and 3b-2 show two representative snapshots of the imbibition of decane+decanoic acid mixture into the calcite nanopore. The imbibition of the mixture oil shows qualitatively new features compared to that of pure oil. First, while a thin precursor film still moves ahead of the oil meniscus, the film is free of decanoic acid. Second, decanoic acid molecules are imbibed into the pore, but they largely remain near the pore entrance during our simulations. Further, the decanoic acid molecules and, in particular, their polar COOH group, are substantially enriched near the calcite wall. For example, as shown in Fig. 4, at $36 \leq t \leq 50$ ns, near the pore entrance (i.e., those within the red dashed box in Fig. 4's inset) the COOH group of decanoic acid approaches closer to the calcite surface than the $CH_2$ groups of decanoic acid and decane molecules and forms a sharp density peak. The enrichment of the COOH group near the calcite relative to the bulk portion of the pore, as indicated by the ratio of the height of the first density peak and the density at the pore center, is more significant than the $CH_2$ groups of decanoic acid and decane molecules.

The much stronger adsorption of the polar COOH motif of decanoic acid on calcite walls than that of the motifs of nonpolar decane leads to the above new features of mixture oil imbibition, and ultimately the slower imbibition of mixture oil and the preferential imbibition of decane over decanoic acid shown in Fig. 2b.

Table 1 shows that the interaction energy of the polar COOH group with the calcite wall is more than 14 times stronger than that of the $CH_2$ motifs of decane and decanoic acid, in line with the observation that polar oil components can interact strongly with mineral surfaces.[18] Therefore, the COOH groups outcompete the $CH_2$ motifs in adsorbing on calcite walls. Further, the stronger association between COOH groups and calcite walls reduces the transport of decanoic acid molecules much more than that of adsorbed decane molecules. For this reason, unlike the decane molecules, even though decanoic acid molecules can reach the tip of the oil meniscus, they do not readily move toward the pore's left end, thereby leading to a precursor film that is essentially free of decanoic acid. Therefore, decanoic acid molecules can only move toward the pore interior via capillary flow but decane molecules can additionally move into the pore through the precursor film, which leads to the selectivity toward the nonpolar decane (i.e., a selectivity factor *S* smaller than 1, see Fig. 2b).

**Table 1**. The interaction energy (in the unit of kJ/mol) of the $CH_2$ and COOH motifs of decane and decanoic acid molecules adsorbed on calcite pore walls during the imbibition process.[†]

|  |  | pure decane | decane-decanoic acid mixture | |
| --- | --- | --- | --- | --- |
|  |  | $CH_2$ motif | $CH_2$ motif | COOH motif |
| dry pore walls | motif- wall | -6.02 ± 0.05 | -5.35 ± 0.36 | -75.92 ± 10.92 |
| prewetted pore walls | motif- wall | -0.79 ± 0.05 | -0.68 ± 0.05 | -11.98 ± 3.72 |
|  | motif- water | -1.61 ± 0.06 | -2.00 ± 0.18 | -53.26 ± 6.74 |

† A motif is adsorbed on dry (prewetted) pore walls if it is within 0.50 nm (0.75 nm) of the topmost atom layer of the wall.

The slower imbibition of the decane+decanoic acid mixture than pure decane can also be traced to the stronger interactions of COOH groups with the calcite walls than nonpolar groups. A large fraction of the decanoic acid molecules imbibed into the pore adsorbs on calcite walls (Fig. 4). These adsorbed molecules have low mobility and accumulate near the pore entrance. Because the adsorbed decanoic acid molecules do not fully displace decane molecules from pore walls but instead intermingle with the interfacial decane molecules (cf. the decane carbon density profiles in Fig. 4), the transport of decane molecules near pore walls is suppressed. Further, the decanoic acid molecules in the middle portion of the narrow pore can interact strongly with those adsorbed on pore walls through their polar groups. The molecular friction associated with such interactions slows down the transport of decanoic acid (and, in turn, decane) through the middle portion of the pore. Because of these effects, the decanoic acid molecules at the pore entrance effectively impede the imbibition of the mixture oil through the entire pore cross-section, thereby leading to slower imbibition compared to the pure decane.

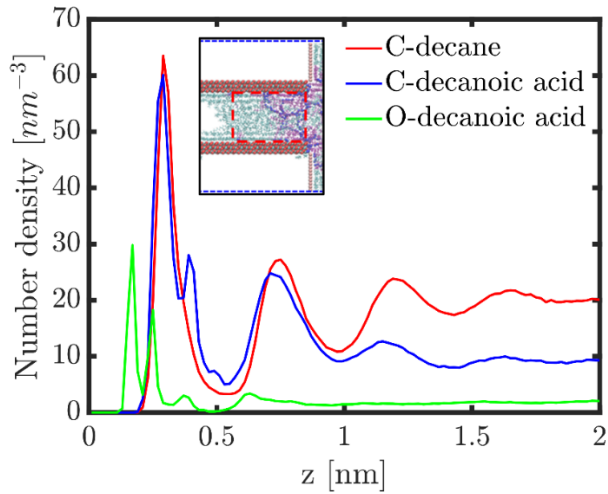

**Figure 4**. The average number density profiles of the carbon and oxygen atoms of the imbibed decane and decanoic acid molecules near the pore entrance (19.03 ≤ $x$ ≤ 24.03 nm; enclosed by the red dashed box in the inset) across the dry nanopore at time 36 ns ≤ $t$ ≤ 50 ns. Only data near the lower wall is shown due to symmetry.

**3.2 Imbibition in prewetted nanopores**

In this section, we investigate the imbibition of pure decane and decane+decanoic acid mixture into calcite nanopores prewetted with a thin water film corresponding to an environment RH of 30% and compare the results with those in Section 3.1.

*Imbibition of decane*. Figure 5a shows that spontaneous imbibition of pure decane occurs even though the pore walls are prewetted by thin water films and decane is clearly hydrophobic. Water films slow down the imbibition of pure decane significantly compared to the dry pore case. Despite that decane is hydrophobic and calcite walls are wetted by water films, capillary flow can still drive decane imbibition into the pores because of the favorable spreading of wetted calcite walls by the decane. To appreciate this, we set up a separate simulation in which a decane droplet is placed on the same prewetted calcite walls used in the imbibition simulations. As shown in Fig. 6, depositing a decane droplet on the calcite substrate prewetted by the molecularly thin water film does not displace the water film. This is because the small and highly polar water molecules interact more strongly with the calcite surface than the nonpolar decane molecules. For the partial wetting of decane on a calcite substrate coated by a thin water film, the calcite substrate and the water film atop can be considered as one entity. The contact angle $\theta$ of decane with a surface tension of $\gamma_o$ on this entity follows $\cos\theta = (\gamma_{wc} - \gamma_{wc-o})/\gamma_o$, where $\gamma_{wc}$ is the surface tension of this entity and $\gamma_{wc-o}$ is the interfacial tension between this entity and liquid decane. The calcite-water film entity features a high-energy surface because of the strong water-water and water-calcite intermolecular interactions. Therefore, similar to high-energy surfaces such as water surfaces, interfacial tension of this entity with nonpolar fluids $\gamma_{wc-o}$ is smaller than its surface tension $\gamma_{wc}$.[41] Consequently, the contact angle $\theta$ of a decane droplet on the calcite surface coated by the thin water layer is less than 90° as shown in Fig. 6.

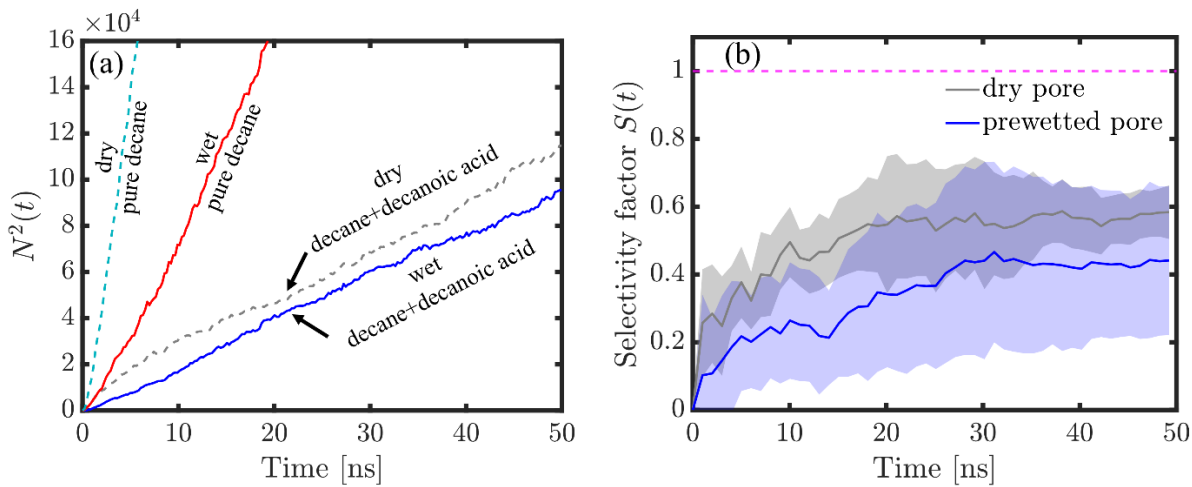

**Figure 5**. (**a**) A comparison of the evolution of the total number of oil molecules (decane + decanoic acid) imbibed into a pre-wetted calcite nanopore (solid line) to that of a dry calcite nanopore (dashed line) as a function of time. (**b**) A comparison of the evolution of the selectivity factor of decanoic acid relative to decane in prewetted and dry calcite nanopore as a function of time. The shade around each curve shows the standard deviation of the selectivity factor.

The partial wetting of calcite substrates with a water film by decane leads to a capillary pressure that drives the imbibition of decane into prewetted calcite nanopores. Such a capillary pressure is smaller than that in dry calcite pores where complete wetting occurs. Therefore, the imbibition of decane into prewetted calcite nanopore is slower than into dry calcite pores (Fig. 5a).

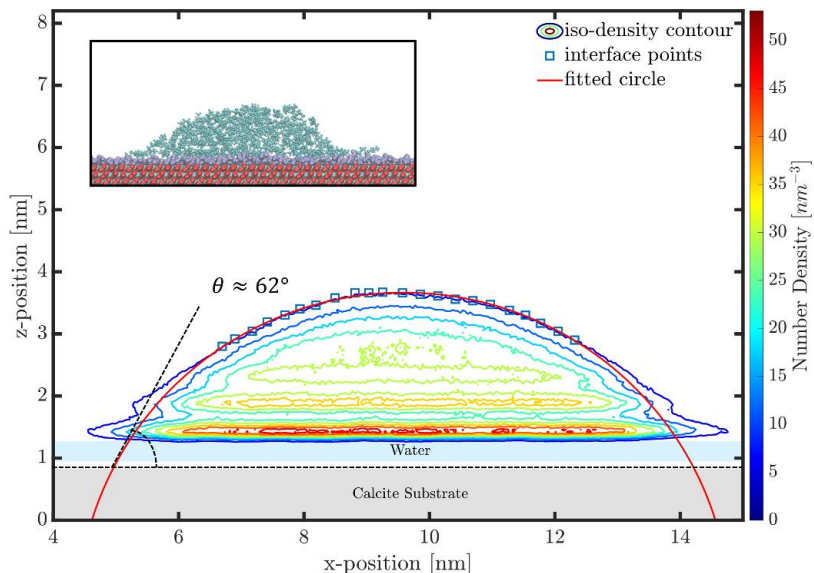

**Figure 6**. The iso-density contour of decane molecules in a decane droplet positioned on a calcite surface prewetted by a thin water film (the water film has the same area density of 11.06 nm$^{-2}$ as in the prewetted pore system in Fig. 7). The red line is a quadratic fit to an iso-density of 5.4 nm$^{-3}$ near the droplet's top surface. The horizontal dashed line denotes the position of the top layer of the calcite wall's oxygen atoms. The inset is a snapshot of the droplet on the prewetted calcite wall.

To further understand how the prewetting of calcite walls by thin water films affects the decane imbibition, we studied the molecular processes of the imbibition. Figures 7a-1 and 7a-2 show two representative snapshots of the imbibition process. A significant difference from those shown in Fig. 3, is the absence of the precursor decane film ahead of the meniscus. The vanished precursor film can be attributed to the weaker interactions between decane molecules and the pore walls. As shown in Table 1, when the thin water film is introduced, the interaction energy between the decane in the first layer and the calcite wall is reduced by 7.7 times. Although such a reduction is partially compensated by the decane-water interactions, it greatly reduces the energetic gain when a decane molecule is transported from bulk decane to a thin decane film on pore walls and contributes to the disappearance of the precursor film. The latter eliminates a transport pathway for the imbibition of decane into the calcite pore and thus contributes to the slower imbibition shown in Fig. 5a.

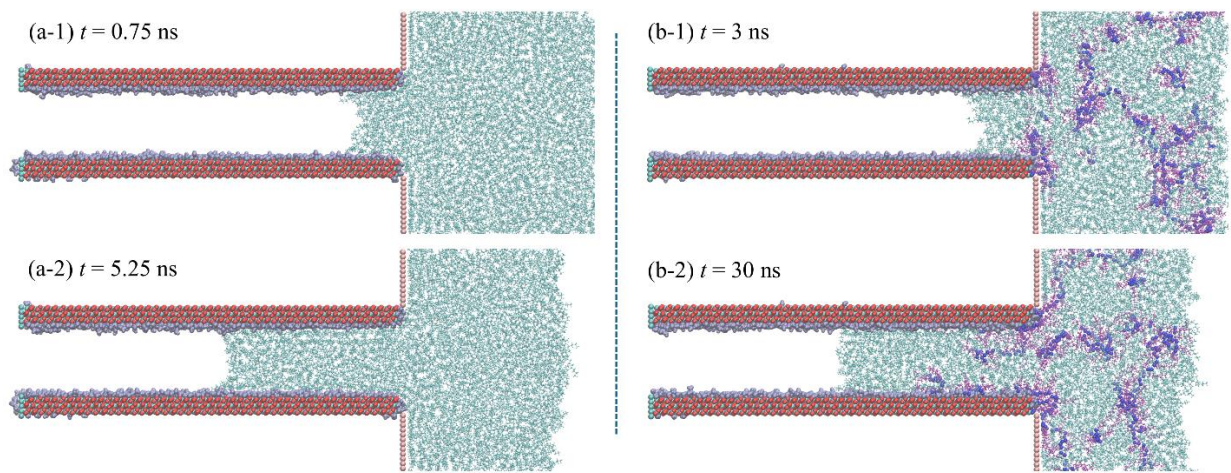

**Figure 7**. Snapshots of the imbibition of pure decane (a) and decane-decanoic acid mixture (b) into a pre-wetted calcite nanopore. The color coding and molecular representation are the same as defined in Fig. 1.

*Imbibition of decane+decanoic acid mixture*. Figures 5a and 5b reveal three key features. First, the imbibition of the decane+decanoic acid mixture into prewetted pores is slower than that of pure decane, but the slowdown is much less significant than in dry pores. For example, Fig. 5a shows that, in prewetted pores, it takes 4 times longer (48 ns *vs*. 12 ns) to imbibe 300 oil molecules from a decane+decanoic acid reservoir than from a pure decane reservoir; in dry pores, it takes 11.8 times longer (40.1 ns *vs*. 3.4 ns). Second, the overall oil imbibition rate in a prewetted pore is rather similar to that in a dry pore. Third, the strong selectivity toward nonpolar decane in the imbibition of decane+decanoic acid mixture is not greatly changed by the thin water films within statistical uncertainty.

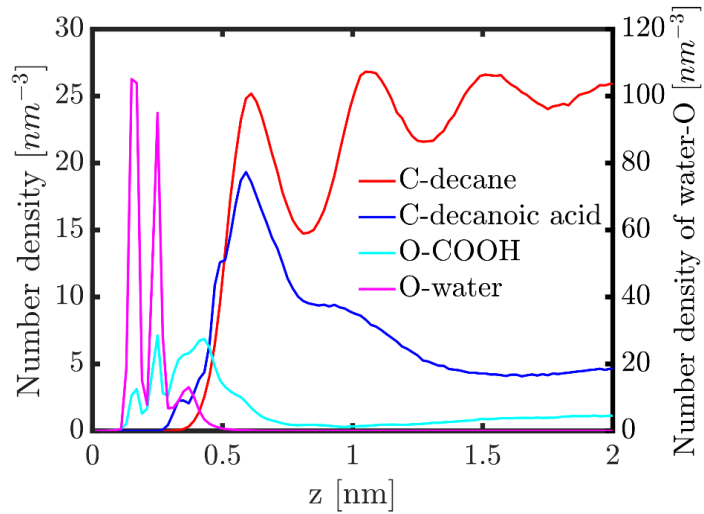

**Figure 8**. The average number density profiles of the carbon and oxygen atoms of the imbibed decane and decanoic acid molecules near the pore entrance (19.09 ≤ $x$ ≤ 24.09 nm) across the prewetted calcite pore at time 12 ns ≤ $t$ ≤ 50 ns. Only data near the lower wall is shown due to symmetry.

The first observation can be traced to the reduced strength of the adsorption of the decanoic acid molecules' COOH groups. As shown in Fig. 8, in the presence of thin water films, the COOH groups are still greatly enriched near the calcite walls. Just as in dry pores, they have lower mobility than decane molecules and thus slow down imbibition compared to the pure decane case. However, the interaction energy between the COOH groups and the calcite wall becomes ~1/6 of that in dry pores (Table 1) and the COOH groups interact strongly with the interfacial water (Table 1). Thus, the COOH groups become hydrated by small, highly mobile water molecules, which can be seen from the density profile of the O atoms of COOH groups and water molecules in Fig. 8. Meanwhile, the contact adsorption of decanoic acid molecules' $CH_2$ motifs are largely eliminated. Indeed, as shown in Fig. 8, nonpolar carbon atoms of decanoic acid molecules form a single peak farther from the wall than the COOH groups, indicating that once their COOH groups are adsorbed on the calcite surface and/or in the water film, their aliphatic tails protrude into the pore interior. Given these changes, the pore-clogging effect by decanoic acid molecules becomes weaker than in the dry pores, and the reduction of oil imbibition in prewetted pores is weaker than in dry pores.

The second observation of the similar overall imbibition rates of decanoic acid+decane mixture in dry and prewetted pores originates from the dual effects of water film on oil transport. On the one hand, the water film tends to enhance oil imbibition by mitigating the pore-clogging effects caused by the strong interactions between polar COOH groups and calcite walls. On the other hand, the water film eliminates the imbibition contributed by the precursor film and reduces the pore width accessible to oil molecules, which favor slower oil imbibition than in dry pores. The competition between these effects makes the oil imbibition relatively insensitive to the presence or absence of water film on the calcite walls.

The last observation, the similar selectivity of imbibition toward nonpolar decane in dry and prewetted pores, provides insights into the dominating mechanism for the imbibition selectivity. First, as discussed above, the stronger association (and thus friction) of adsorbed decanoic acid molecules with pore walls than interfacial decane molecules contributes to the selectivity toward decane. Second, the transport of decanoic acid molecules into both pores relies partly on capillary flow, and thus the enrichment of decanoic acid (decane) molecules in the interfacial (bulk) region and the weaker flow near pore walls is the second factor contributing to the observed selectivity. The association and friction of adsorbed decanoic acid molecules is weaker near prewetted walls as indicated by their higher mobility there. Therefore, if the first factor dominates the selectivity in the prewetted pores, the selectivity toward nonpolar decane there should be weaker than in dry pores, which differs from the data in Fig. 5b. Therefore, the second factor appears to dominate the selectivity toward nonpolar decane in the prewetted pore.

## 4. Conclusions

In summary, we investigated the imbibition of two model crude oils (pure decane and decane + decanoic acid mixture) into 4 nm-wide calcite pores whose walls are initially dry or wetted by a molecularly thin water film. We show that the imbibition into dry pores is contributed both by capillary flow and propagation of precursor films consisting of only decane. Imbibition of mixture oil is not only slower but also accompanied by a distinct selectivity toward the nonpolar decane. Oil is still imbibed spontaneously into prewetted pores, albeit the rate is slower than in dry pores, especially for pure decane. Thin water films on pore walls eliminate precursor films ahead of the liquid meniscus but have limited effect on the selectivity of imbibition. The latter suggests that the relative enrichment of decanoic acid molecules and weaker flow near pore walls, in the presence or absence of water, dominate the imbibition selectivity in the parameter space we studied.

Impurities are ubiquitous in natural and engineering systems but are often neglected in previous imbibition studies. Our study revealed that impurities, either in fluids and/or on pore walls (e.g., decanoic acid and molecularly thin water film in this work), can affect not only the imbibition rate but also the nature of imbibition (e.g., selectivity toward a particular fluid species). Further, the effects of impurity on the imbibition depend heavily on the molecular interactions between the impurity molecules and pore walls. Therefore, more attention should be paid to impurities and their chemical nature in future imbibition studies. While this necessarily complicates those studies, the insight gained from them may open new avenues for manipulating imbibition and improving consistency of results through intentional control of impurities in experimental studies.

## Acknowledgment

The authors thank the ARC at Virginia Tech for the generous allocation of computing time. R.Q. acknowledges partial support from the NSF under grant number CBET-2246274.

## Data Availability Statement

The data that support the findings of this study are available within the article and its supplementary material.